\documentclass[notitlepage,onecolumn,showpacs,10pt]{revtex4-1}
\usepackage[utf8]{inputenc}
\usepackage{amsmath}
\usepackage{amsfonts}
\usepackage{amssymb}
\usepackage{fullpage}
\usepackage{array,bm}
\usepackage{graphicx,tikz}
\usepackage{graphicx}
\usepackage{bm}
\usepackage[english]{babel}
\usepackage{subcaption}
\usepackage{soul}
\usepackage[hidelinks]{hyperref}

\begin{document}

\title{ Two and three-state quantum heat engines with stochastic resetting }
\author{Ashutosh Kumar}
\email{ashutoshkumarr06@gmail.com}

\author{Sourabh Lahiri} 
\email{sourabhlahiri@bitmesra.ac.in}
\affiliation{Department of Physics, Birla Institute of Technology Mesra, Ranchi, Jharkhand 835215, India}

 \author{Trilochan Bagarti}
\email{trilochan.bagarti@tatasteel.com}
 \affiliation{ Graphene Centre, Tata Steel Limited, Jamshedpur, 831007, Jharkhand, India}

\author{Subhashish Banerjee}
\email{subhashish@iitj.ac.in}
\affiliation{ Indian Institute of Technology, Jodhpur, 342030, Rajasthan, India}

\begin{abstract}
Quantum heat engines have undergone extensive studies over the last two decades. Simultaneously, the studies of the applications of stochastic resetting in various fields are on the rise. We explore the effect of stochastic resetting on the dynamics of a two-level and a three-level quantum heat engine. The extracted work is shown to increase with the resetting rate. 
The effective efficiency that takes into account the work done due to resetting remains constant. 
However, if the work done due to resetting is ignored, then the system can incorrectly imply a  different behaviour, including the false inference that it is not working as an engine at all. 
The efficient power is observed to increase beyond that obtained in the absence of resetting, and is shown to be higher for a three-level engine.
\end{abstract}


\maketitle

\section{Introduction}

Nature has endowed self-sustaining multi-cellular organisms, including ourselves, with myriads of microscopic machines, known as molecular motors. These tiny machines are indispensable in the delivery of food within human cells, muscle contractions, generating forces for self-propulsion, RNA transcription, DNA replication, etc.\cite{mavroidis2004molecular,sperry2007molecular}.
Their artificial counterparts, namely nanorobots, although still in their nascent stage, have fueled studies on microscopic heat engines and refrigerators. The theory as well as experimental studies of these engines have experienced rapid progress in recent years \cite{rossnagel2014nanoscale,blickle2012realization,seifert2012stochastic,Roldan2016}, driven by their potential applications across multiple industries, particularly healthcare. These machines have been proposed for applications in targeted drug delivery, precision surgery, and other advanced medical applications\cite{manjunath2014promising,saadeh2014nanorobotic}. Energy harvesting at mesoscopic scales to power small-scale machines has thus attracted a lot of attention \cite{Phillips2021}.

Quantum thermodynamics is the study of thermodynamical processes from a quantum mechanical perspective \cite{gemmer2009quantum,sekimoto2010physics,alicki2019introduction}. Due to advancements in theory and experiments,  several thermodynamic devices, such as quantum heat engines \cite{thomas2018thermodynamics,kumar2023thermodynamics} and quantum batteries \cite{binder2015quantacell,alicki2013entanglement,tiwari2023impact,bhanja2024impact} have been studied in recent times. Extensive investigations have been made on the development and enhancement of quantum thermal machines, with many studies yielding some surprising results \cite{scully2002quantum,quan2007quantum,quan2005quantum,rostovtsev2003improving}. A quantum heat engine using a 3-level maser was proposed in \cite{scovil1959three}. With the advancement of quantum information technologies \cite{buluta2009quantum,buluta2011natural} and the ability to control single atoms, the field has become very active.
 Notably, studies have explored systems such as a single ion trapped in a linear Paul trap \cite{rossnagel2016single} and ultra-cold atoms \cite{barontini2019ultra}, demonstrating the potential for extracting work from these quantum systems. Additionally, spin-based quantum heat engines have been experimentally realized using nuclear magnetic resonance techniques \cite{peterson2019experimental}. The development of quantum heat engines utilizing one- and two-qubit systems, as well as harmonic oscillators coupled to a squeezed thermal bath, has been investigated \cite{kumar2023thermodynamics,rossnagel2014nanoscale}.

In recent times, the stochastic resetting mechanism has been employed to interrupt the evolution of Brownian motion by returning it to a predetermined position \cite{evans2011diffusion,gupta2022stochastic,pal2015diffusion}. Its effect on the thermodynamics of classical stochastic heat engines has been studied in \cite{lahiri2024efficiency}. An experimental realization of a classical reset protocol using lasers and the associated energy cost was discussed in \cite{Pal2020}.  Stochastic resetting has been used in a quantum setting for a system evolving unitarily \cite{Majumdar2018,Kulkarni2023}. The steady-state density matrix has been shown to be off-diagonal \cite{Majumdar2018}. In \cite{Kulkarni2023} the authors discuss the generation of quantum entanglement by the resetting process. Its implementation on a tight-binding chain has been studied in \cite{Dattagupta2022}. 
Resetting the dynamics of a quantum particle for improving the first passage time \cite{Barkai2023} and modification in uncertainty principle due to resets \cite{barkai2024} have been explored. Resets in quantum many-body systems have been discussed in \cite{Perfetto2021,Perfetto2022}. A generalized Lindblad dynamics in presence of resets has been obtained in \cite{Perfetto2022a}.

In the current work, we apply stochastic resetting to a quantum Otto heat engine in the isochoric steps, where the system is in contact with the respective heat baths. In-between two consecutive resets, the system undergoes Lindbladian evolution. We study the evolution of two-level and three-level systems undergoing this process, and compare their thermodynamics with the no-reset case. The ground and the second excited levels of the three-level system are separated by the same energy gap as that in the two-level system, with the first excited level appearing midway. The work output,  work input for sustaining reset operations, effective efficiency, and the efficient power of the two- and three-level systems are compared. 

We also specify some areas where this study is likely to find applications. Qudits are higher dimensional generalizations of qubits that are increasingly gaining importance in several
areas of quantum physics and technology \cite{adepoju2017joint,dutta2023qudit}. Three-level systems are examples of qudits and are required in a number of applications, such as, the laser \cite{sargent1974laser}. Also, a two-qubit system in the regime where the distance between the qubits is smaller than the environmental length scale becomes effectively three-level \cite{ficek2002entangled}.

The plan of the work is as follows. In Sec. \ref{sec:theory}, we discuss the theoretical formulation of the model, and the effect of stochastic energy resetting on the system state. The section also describes the engine protocol, in particular, the Otto engine, and the definitions of various thermodynamic observables. In Sec. \ref{sec:results}, the output works, resetting works, effective efficiencies, and efficient powers of two-level and three-level engines are studied and compared. 
Section \ref{sec:thermodynamic_reset_work} discusses how to avoid a possible source of confusion regarding the definition of effective heat.
We summarize and conclude in Sec. \ref{sec:conclusions}.

\section{Theory} \label{sec:theory}

\subsection{Evolution without resetting}
Consider a quantum system that is not in contact with any thermal reservoir during its evolution with time $t$, so that it undergoes unitary evolution under the system Hamiltonian $H(t)$. The unitary evolution is given by the von Neumann equation for the density operator $\rho(t)$:
\begin{equation}
    \frac{\partial\rho}{\partial t} = -\frac{i}{\hbar}[H(t),\rho(t)],
    \label{eq:von Neumann}
\end{equation}
where $H(t)$ and $\rho(t)$ are the associated Hamiltonian and density operators, respectively. Throughout this work, we assume that the unitary evolution follows the conditions of quantum adiabaticity.

\begin{figure}[!ht]
    \centering
  %  \hfill
    \begin{subfigure}{0.45\linewidth}
        \centering
        \includegraphics[width=0.45\textwidth]{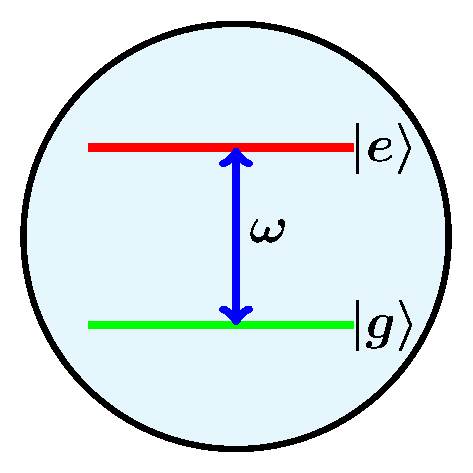}
        \hfill
        \includegraphics[width=0.45\textwidth]{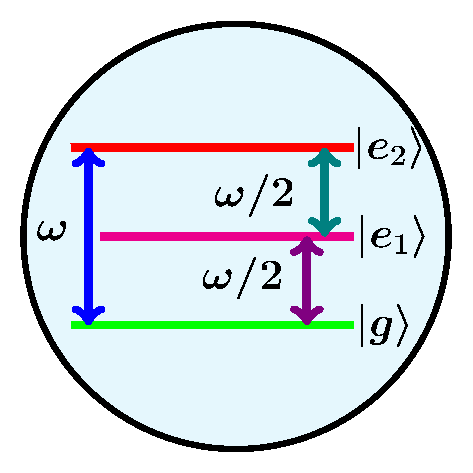}
        \caption{}
    \end{subfigure}
    \hfill
        \begin{subfigure}{0.45\linewidth}
        \centering
        \includegraphics[width=\textwidth]{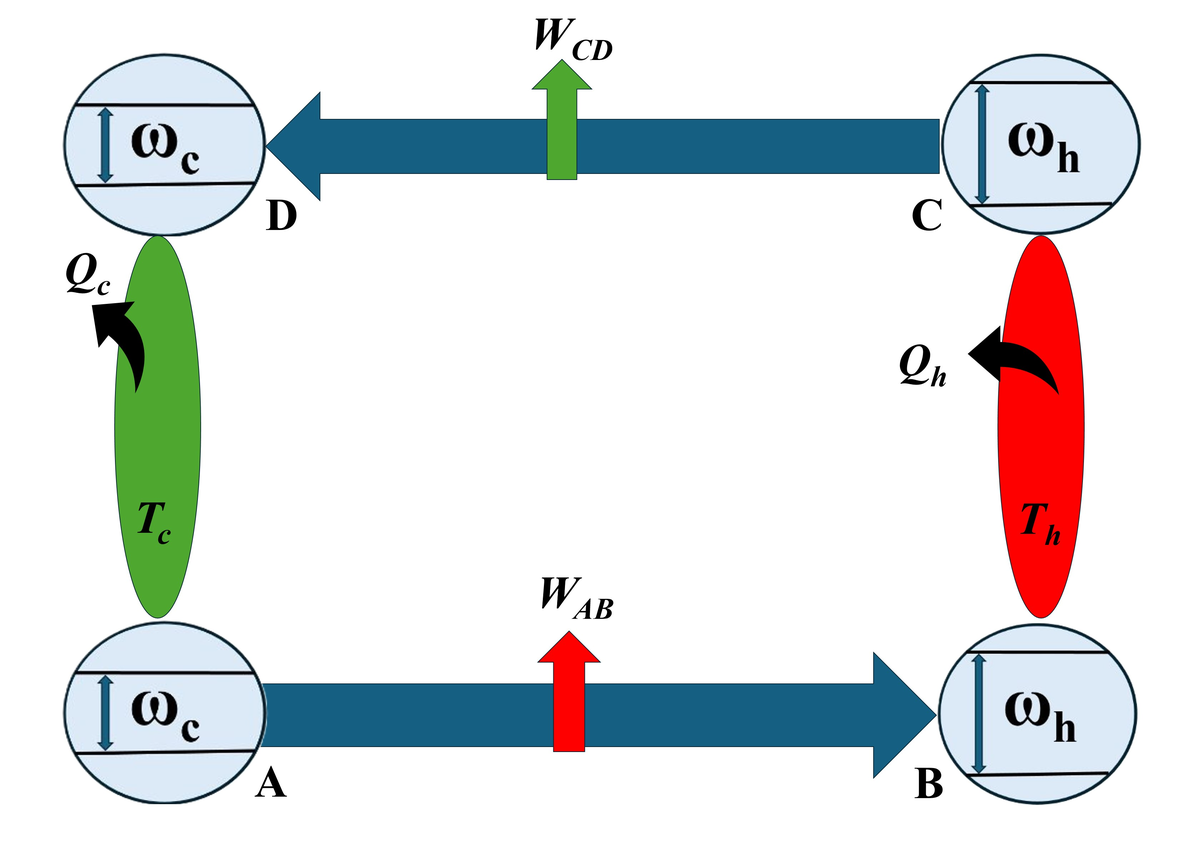}
        \caption{}
    \end{subfigure}
    \caption{
    (a) A schematic diagram of the two- and three-level systems. The two-level system has an energy gap $\omega$ between the states $|e\rangle$ and $|g\rangle$. In the three-level system, $\omega$ represents the energy difference between the states $|e_2\rangle$ and $|g\rangle$, with a gap of $\omega/2$ between consecutive energy levels. (b) A schematic diagram of the Otto engine protocol implemented on the two-level system (see sec. \ref{sec:thermodynamics}).}
    \label{fig:states}
\end{figure}

When the system is in contact with a thermal reservoir at temperature $T$, it follows the Lindblad dynamics \cite{breuer2002theory,banerjee2018open,gardiner2004quantum}, given by
\begin{equation}
    \frac{\partial\rho}{\partial t} = -\frac{i}{\hbar}[H(t),\rho(t)] + \sum_i \gamma_i (L_i \rho(t) L_i^\dagger - \{L_i^\dagger L_i,\rho(t)\}/2) \equiv \mathcal{L}^{\mathrm{nr}}[\rho(t)].
    \label{eq:Lindblad}
\end{equation}
Here, $L_i$ are the ladder operators and $\gamma_i$ are the damping rates. The superoperator $\mathcal{L}^{\mathrm{nr}}$ evolves the system as per the Lindblad equation, in absence of resetting. 
The underlying assumptions are the  weak coupling between the system and the heat bath (Born approximation), memoryless evolution and fast relaxation of heat bath (Markov approximation), and the rotating wave approximation \cite{breuer2002theory}. 
%%%%
\paragraph*{\textbf{Two-level system:}} For a two-state system, $L_1 = \sigma_-$ and $L_2=\sigma_+$ (these ladder operators are defined in terms of the Pauli matrices by $\sigma_{\pm} = \sigma_x \pm i\sigma_y$), while $\gamma_1 = \gamma_0 (n_{\mathrm{th}}(\omega)+1)$ and $\gamma_2 = \gamma_0 n_{\mathrm{th}}(\omega)$. Here, $\gamma_1$ and $\gamma_2$ give the rate of transition from ground to excited state and vice versa, respectively. 
The parameter $\gamma_0 = 4\omega^3 |\hat{d}|^2/(3\hbar c^3)$ represents the spontaneous emission rate, where $\omega$ is the energy difference between two states, $\hat{d}$ is the dipole moment operator, $\hbar$  is the reduced Planck constant, and  $c$ is the speed of light. We set $4|\hat{d}|^2/(3 c^3)=\hbar=1$ throughout this article, so that $\gamma_0 = \omega^3$. If the inverse temperature of the bath is $\beta=1/k_B T$, where $k_B$ is the Boltzmann constant, then $n_{\mathrm{th}}(\omega)$ is given by
$n_{\mathrm{th}}(\omega) = 1/(e^{\beta\omega}-1)$.
The Hamiltonian $H(t)$ is of the form ($\sigma_z$ is the third Pauli matrix)
\begin{equation}
    H(t) = \frac{1}{2}\hbar\omega(t)\sigma_z =
    \frac{1}{2}
    \hbar\omega(t)
    \left(
    \begin{array}{cc}
        1 & 0\\
        0 & -1
        \end{array}\right).
    \label{eq:H_twostate}
\end{equation}
\paragraph*{\textbf{Three-level system:}} For a three-state system, the ladder operators are given by \cite{zettili2009quantum}
\begin{equation}
    S_+ = \sqrt{2}
    \left(
    \begin{array}{ccc}
        0 & 1 & 0\\
        0 & 0 & 1\\
        0 & 0 & 0
        \end{array}
    \right); \hspace{0.5cm}
    S_- = \sqrt{2}
    \left(\begin{array}{ccc}
        0 & 0 & 0\\
        1 & 0 & 0\\
        0 & 1 & 0
    \end{array}\right).
\end{equation}
The Hamiltonian is of the form 
\begin{equation}
    H(t) =  \frac{1}{2}
    \hbar\omega(t)
    \left(\begin{array}{ccc}
        1 & 0 & 0\\
        0 & 0 & 0\\
        0 & 0 & -1
    \end{array}\right).
\end{equation}
In this case, Eq. (\ref{eq:Lindblad}) becomes (see the general formalism in \cite{breuer2002theory})
\begin{eqnarray}
    \frac{\partial\rho}{\partial t} = -\frac{i}{\hbar}[H(t),\rho(t)] &+& \gamma_{\mathrm{em}}^{(1)}(N(\omega/2)+1)\big[S_- \rho(t)S_+ - \{S_+ S_-,\rho(t)\}/2\big]\nonumber\\
    &+& \gamma_{\mathrm{ab}}^{(1)}N(\omega/2)\big[S_+ \rho(t)S_- - \{S_- S_+,\rho(t)\}/2\big] \nonumber\\
    &+& \gamma_{\mathrm{em}}^{(2)}(N(\omega)+1)\big[S_-^2 \rho(t)S_+^2 - \{S_+^2 S_-^2,\rho(t)\}/2\big]\nonumber\\
    &+& \gamma_{\mathrm{ab}}^{(2)}N(\omega)\big[S_+^2 \rho(t)S_-^2 - \{S_-^2 S_+^2,\rho(t)\}/2\big].
    \label{eq:Lindblad_3L}
\end{eqnarray}
The rates $\gamma_{\mathrm{em}}^{(1)}$ and $\gamma_{\mathrm{ab}}^{(1)}$ are the damping rates associated with the downward (emission) and upward (absorption) transitions between consecutive levels, i.e., between levels 1 and 2, or 2 and 3. Similarly, $\gamma_{\mathrm{em}}^{(2)}$ and $\gamma_{\mathrm{ab}}^{(2)}$ are the transition rates between levels 1 and 3. 
The energy eigenvalues for the two- and three-level systems are schematically shown in Fig. \ref{fig:states}(a).

\subsection{Evolution with resetting}{\label{resetting Evolution}}

We now apply the reset operation to a quantum system in contact with a heat bath. Under this operation, the instantaneous density operator of the system is very quickly (compared to the equilibration time) brought to a pre-decided constant density operator, like $\rho_g$, corresponding to the particle being reset to its ground state. The reset events take place with a given rate $r$ that can be experimentally controlled. It can be readily checked that this definition leads to an exponential distribution of the time interval $\tau$ between consecutive reset events \cite{gupta2022stochastic} :
\begin{equation}
    p(\tau) = r e^{-r\tau}.
    \label{eq:tau distribution}
\end{equation}
In presence of resetting, the dynamics of the system evolves as follows:
\begin{equation}
  \rho^{\mathrm{r}}(t+dt) =
  \left\{
    \begin{array}{ll}
        \rho^{\mathrm{reset}} ~\hbox{with probability}~ rdt,\\
        \rho^{\mathrm{r}}(t) + \mathcal{L}^{{\mathrm{nr}}}[\rho^{\mathrm{r}}(t)]dt, ~\hbox{with probability}~ (1-rdt).
    \end{array}
    \right.
    \label{eq:reset dynamics}
\end{equation} 
The above equation says that in the time interval $(t,t+dt)$, the evolving state (as per master equation) can reset to the chosen state $\rho^{\mathrm{reset}}$ with probability $rdt$, or can continue to evolve as per the Lindblad master equation with probability $(1-rdt)$. $\rho^{\mathrm{r}}(t)$ is the density operator at time $t$, the superscript indicating a finite resetting rate. The superscript ``${\mathrm{nr}}$'' indicates the usual Lindbladian evolution, i.e, when there is no resetting. 

 Let $\tilde\rho^{\mathrm{nr}}(t_f;t_i)$ be the solution of the Lindbladian evolution $\mathcal{L}^{\mathrm{nr}}$ from time $t_i$ to time $t_f$. Let $\rho^{\mathrm{nr}}(t)\equiv \tilde\rho^{\mathrm{nr}}(t;0)$ be the solution of $\mathcal{L}^{\mathrm{nr}}$ when $t_i=0$ and $t_f=t$.
The density operator $\rho^{\mathrm{r}}(t)$ is related to $\rho^{\mathrm{nr}}(t)$ via the renewal equation \cite{gupta2022stochastic,lahiri2024efficiency,Majumdar2018,Perfetto2022a}:
\begin{equation}
    \rho^{\mathrm{r}}(t) = \rho^{\mathrm{nr}}(t) e^{-rt} + r\int_0^t d\tau ~\tilde\rho^{\mathrm{nr}}(t;t-\tau) e^{-r\tau}.
    \label{eq:renewal equation}
\end{equation}
The two terms on the right-hand-side of Eq. (\ref{eq:renewal equation}) correspond to the two cases mentioned in Eq. (\ref{eq:reset dynamics}). 
The Lindbladian evolution from time 0 to $t$ takes place with a probability $e^{-rt}$. This is the probability of obtaining the density operator $\rho^{nr}(t)$. This contribution is given by the first term. The second term, on the other hand, provides the contribution coming from the reset events. 
The tilde symbol over the density matrix appearing inside the integral implies that it is the solution to the Lindblad equation with the initial state being the last reset state. This is different from $\rho^{\mathrm{nr}}(t)$ appearing in the first term, where the initial state is the global initial state at the beginning of the isochoric stroke. Consider the integrand: 
\[rd\tau~e^{-r\tau}\tilde\rho^{\mathrm{nr}}(t;t-\tau).\] 
The probability of the system being reset within the time step $(t-\tau-d\tau,~ t-\tau)$ is given by the factor $rd\tau$. 
The factor $\exp(-r\tau)$ gives the probability that since time $t-\tau$, a  Lindbladian evolution  has taken place till  time $t$. 
Thus, the product $[rd\tau~\exp(-r\tau)]$ is the probability of the last reset taking place at time $t-\tau$, \textit{and} a subsequent Lindbladian evolution for a duration $\tau$. This is the probability of obtaining the density operator $\tilde\rho^{\mathrm{nr}}(t;t-\tau)$. Since the last reset can take place in \textit{any} time step in the interval $(0,t)$, the integral sums over all such possible contributions.
Note that only the last reset is of consequence here, since the memory of the entire history gets erased at every reset event.

Thus, the theory of Lindbladian evolution under stochastic resetting can be entirely formulated in terms of states undergoing pure Lindbladian evolution, by means of the above renewal equation.

\subsection{The Model and its Thermodynamics}
\label{sec:thermodynamics}

We briefly explore the operation of a quantum heat engine that works on the principles of the Otto Cycle where there is a clear separation between the steps involving transfer of energy either as work or as heat. The cycle comprises four distinct thermodynamic strokes. There are two adiabatic processes, where the quantum system is isolated from thermal interaction, and two isochoric processes, during which the system exchanges energy with thermal reservoirs at different temperatures. 
The schematic diagram showing the evolution of the two-level system under the Otto engine protocol is shown in Fig. \ref{fig:states}(b).

The work \textit{done on} and the heat \textit{absorbed by} the system are given by the difference between the final and the initial average energies of the system during the respective strokes. In case these values are negative, it will be understood that the work is \textit{extracted from}  or heat is \textit{dissipated by} the system, respectively.   Only during the isochoric strokes, the reset dynamics (see Sec. \ref{resetting Evolution}) is applied. The details of the four strokes are given below.
\begin{enumerate}
      \item \textbf{Adiabatic expansion (without reset), stroke $A\to B$:} This is the stroke that takes the system from state $A$ (described by the density operator $\rho_A$) to a state $B$ (described by the density operator $\rho_B$). The system evolves under dynamics governed by the von Neumann equation, given by  Eq. (\ref{eq:von Neumann}), from time $t=0$ to $t=\mathcal{T}$. The energy level spacings of the system increase linearly with time, from $\omega_c$ to $\omega_h$. The Hamiltonian of the system changes from $H(\omega_c)$ to $H(\omega_h)$. The functional form of $\omega(t)$ is
    \begin{equation}
    \label{eq:H_exp_com}
    \omega(t) = \omega_c \left(1 - \frac{t}{\mathcal{T}}\right) + \omega_h \frac{t}{\mathcal{T}}.
    \end{equation}

Since the system is not connected to any heat bath during this stroke, the associated change in mean energy yields the work done on the system:
\begin{equation}
    W_{A \to B}= \langle E_B\rangle - \langle E_A\rangle =  \mathrm{Tr}[\rho_{B}H(\omega_{h})]-\mathrm{Tr}[\rho_{A}H(\omega_{c})].
     \label{eq:Wab}
\end{equation}
Here, the angular brackets denote averaging with respect to the corresponding density operator. It may be noted that if the system is running in the engine mode, $W_{A\to B}$ will be negative, implying that work is extracted in this stroke.

    \item \textbf{Isochoric heating (with reset), stroke $B\to C$:} This stroke takes place from time $t=\mathcal{T}$ (state $B$, described by $\rho_B$) to time $t=2\mathcal{T}$ (state $C$, described by $\rho_C$). In this stroke, the system is connected with the hot bath at temperature $T_h$, and the Hamiltonian is held constant at $H(\omega_h)$. The evolution is interrupted by resetting the system at the given rate to the chosen reset state. For the two-level system, this is either the ground state $|g\rangle$ or the excited state $|e\rangle$. For the three-level system, it can be any state among $|g\rangle$ (ground state), $|e_1\rangle$ (first excited state), or $|e_2\rangle$ (second excited state). The detailed dynamics of the resetting process are discussed in Section (\ref{resetting Evolution}).
    During this stroke, there being no change in the Hamiltonian, the change in average energy gives the effective heat \textit{absorbed by} the system. This is given by the sum of two terms: $Q_{B\to C}$ and $W_{h,\mathrm{reset}}$. The former term is the sum of heat energies absorbed by the system when the evolution is Lindbladian, i.e., in the time-intervals between consecutive reset events. The latter term is the total energy input required to bring about all the reset events during this stroke. We can then write the effective absorbed heat as
\begin{equation}
    \tilde{Q}_{B\to C} = \langle E_C\rangle - \langle E_B\rangle =  \mathrm{Tr}[\rho_{C} H(\omega_{h})] - \mathrm{Tr}[\rho_{B} H(\omega_{h})] = Q_{B\to C} + W_{h,\mathrm{reset}}.
    \label{eq:Q_BC}
\end{equation}
The fact that $W_{h,{\mathrm{reset}}}$ is already a part of $\tilde Q_{B\to C}$ can be explained as follows. Suppose the resets take place at times $t_1,~t_2,~\cdots,~t_N$, where $t_1 \leq t_2 \leq \cdots \leq t_N$. Then the net change in energy can be split up as
\begin{eqnarray}
    \langle E_C\rangle - \langle E_B\rangle &=& \left[\langle E(t_1^-)\rangle - \langle E_B\rangle\right] + \left[\langle E(t_1^+)\rangle - \langle E(t_1^-)\rangle\right] + \left[\langle E(t_2^-)\rangle - \langle E(t_1^+)\rangle\right] \nonumber\\
    &+& \left[\langle E(t_2^+)\rangle - \langle E(t_2^-)\rangle\right] + \cdots + \left[\langle E_C\rangle - \langle E(t_N^+)\rangle\right]\nonumber\\
    &=& \sum_{i=0}^{N} \left[\langle E(t_{i+1}^-) \rangle - \langle E(t_i^+) \rangle \right] + \sum_{i=1}^N \left[\langle E(t_{i}^+) \rangle - \langle E(t_i^-) \rangle \right],
\end{eqnarray}
where $\langle E(t_0) \rangle = \langle E_B \rangle$, $\langle E(t_{N+1}) \rangle = \langle E_C \rangle$.
The first term is $Q_{B\to C}$ and the second term is $W_{h,{\mathrm{reset}}}$.
\item \textbf{Adiabatic compression (without reset), stroke $C\to D$:} This takes place from time $t=2\mathcal{T}$ to $t=3\mathcal{T}$, and is essentially the reverse of the process $A\to B$. The system state evolves unitarily from $\rho_C$ to $\rho_D$, with $\omega(t)$ changing linearly from $\omega_h$ to $\omega_c$. The dynamics is given by the von Neumann equation. The explicit form of $\omega(t)$ is 
\begin{equation}
    \omega(t) = \omega_c \left(\frac{t}{\mathcal{T}}-2\right) + \omega_h\left(3- \frac{t}{\mathcal{T}}\right).
\end{equation}
The work done on the system during this stroke is given by
\begin{equation}
    W_{C \to D} = \langle E_D\rangle - \langle E_C\rangle = \mathrm{Tr}[\rho_{D}H(\omega_{C})]-\mathrm{Tr}[\rho_{C}H(\omega_{H})].
    \label{eq:Wcd}
\end{equation}

\item \textbf{Isochoric cooling (with reset), stroke $D\to A$:} This is the final stroke that completes the cycle, taking place from time $t=3\mathcal{T}$ to $t=4\mathcal{T}$. The state of the system evolves from $\rho_D$ to $\rho'_A$, where $\rho'_A = \rho_A$ only when the time-periodic steady state (TPSS) has been reached, which is generally a non-equilibrium one unless the process is quasi-static $(\mathcal{T}\gg 1/\omega)$. The system is connected to a cold bath at temperature $T_c$ in this stroke, and is allowed to evolve with a fixed energy spacing $\omega_c$. However, the dynamics is interrupted by stochastic resetting to a predefined reset state at the given rate. The predefined reset state in this stroke may or may not be the same as that during the stroke $B\to C$. The effective heat absorbed during this stroke is (assuming a TPSS).
\begin{equation}
    \tilde{Q}_{D\to A} = \langle E_A\rangle - \langle E_D\rangle =  \mathrm{Tr}[\rho_{A} H(\omega_{c})] - \mathrm{Tr}[\rho_{D} H(\omega_{c})] = Q_{D\to A} + W_{c,\mathrm{reset}},
    \label{eq:Q_DA}
\end{equation}
where the terms are defined in a similar way as in Eq. (\ref{eq:Q_BC}). Note that the net change in the energy of the system is caused by energy absorbed from both the work source (resetting device) and the heat source (thermal bath).

\end{enumerate}

\paragraph*{\textbf{Efficiency:}}  The effective efficiency $\eta_{\mathrm{eff}}$ under the effect of stochastic resets can be defined as
\begin{equation}
    \eta_{\mathrm{eff}} \equiv -\frac{W_{A\to B} + W_{C\to D}}{\tilde{Q}_{B\to C}},
    \label{eq:eta}
\end{equation}
where the minus sign is added because the output work is negative of the input work:
\[
W_{\mathrm{out}} = -(W_{A\to B} + W_{C\to D}) = -W.
\]
However, if we ignore the energy inputs coming from reset events, the (incorrect) efficiency becomes
\begin{equation}
    \eta \equiv -\frac{W_{A\to B} + W_{C\to D}}{Q_{B\to C}} = -\frac{W_{A\to B} + W_{C\to D}}{\tilde{Q}_{B\to C} - W_{h,{\mathrm{reset}}}},
    \label{eq:eta effective}
\end{equation}
where we have used Eq. (\ref{eq:Q_BC}) to write $Q_{B\to C} = \tilde{Q}_{B\to C} - W_{h,{\mathrm{reset}}}$.
The reset work $W_{h,\mathrm{reset}}$ can be obtained as follows. Suppose the two-level system in the $B\to C$ stroke is reset to the excited level and for the $D\to A$ stroke to the ground level (see Fig. \ref{fig:work_output_vs_rr_diff_reset_state} and the associated discussion regarding the choice of reset states). Then for a single reset event during the $B\to C$ stroke, the energy change is given by $\Delta E = +\hbar\omega_h$ if just before this event the system was in its ground state, otherwise $\Delta E = 0$. In a single such event in the $D\to A$ stroke, $\Delta E = -\hbar\omega_c$ if the state just before the reset was an excited state, otherwise $\Delta E = 0$. If all such reset events are recorded during the simulations, then the resultant input work can be readily computed.
This can be done due to the absence of off-diagonal terms in our density matrix throughout the evolution, as can be checked by explicitly evaluating the right-hand side of Eqs. (\ref{eq:Lindblad}) and (\ref{eq:Lindblad_3L}) by using the matrix forms of the operators involved.
We will find that for the set of parameters used in our case, $\eta$ turns out to be negative, even though $\eta_{\mathrm{eff}}$ remains positive and the system perfectly works as an engine. In other words, if there is a heat source and a work source that simultanesouly provide input energy to the system, then ignoring the work source (or vice versa) can lead to a drastically incorrect characterization of the thermodynamic properties of the system.

The analytical expressions to compute $W_{\mathrm{reset}}^{(2L)}$ for the two-level system, during the $B\to C$ and $D\to A$ strokes are respectively
\begin{eqnarray}
    W_{h,{\mathrm{reset}}}^{(2L)} &=& r\hbar\omega_h\int_\mathcal{T}^{2\mathcal{T}} dt~\rho_{gg}^{\mathrm{r}}(t); \nonumber\\
    W_{c,{\mathrm{reset}}}^{(2L)} &=& - r\hbar\omega_c\int_{3\mathcal{T}}^{4\mathcal{T}} dt~\rho_{ee}^{\mathrm{r}}(t),
    \label{eq:W reset}
\end{eqnarray}
where $\rho_{ee}^{\mathrm{r}}$ and $\rho_{gg}^{\mathrm{r}}$ are the population densities of the excited and the ground states, respectively. 
For the three-level system, we consider the reset state to be the second excited state in the $B\to C$ stroke and the ground state in the $D\to A$ stroke. The normalized populations in the ground, first excited and second excited levels are given by $\rho_{gg}^{\mathrm{r}}$, $\rho_{e_1 e_1}^{\mathrm{r}}$ and $\rho_{e_2 e_2}^{\mathrm{r}}$, respectively. We can then write (with the superscript $(3L)$ indicating the three-level system)
\begin{eqnarray}
    W_{h,{\mathrm{reset}}}^{(3L)} &= r\hbar\omega_h\int_\mathcal{T}^{2\mathcal{T}} dt~\left(\rho_{gg}^{\mathrm{r}}(t)+\frac{1}{2}\rho_{e_1 e_1}^{\mathrm{r}}(t)\right);\nonumber\\
    W_{c,{\mathrm{reset}}}^{(3L)} &=- r\hbar\omega_c\int_{3\mathcal{T}}^{4\mathcal{T}} dt~\left(\rho_{e_2 e_2}^{\mathrm{r}}(t) + \frac{1}{2}\rho_{e_1 e_1}^{\mathrm{r}}(t)\right),
    \label{eq:W reset 3 state}
\end{eqnarray}
where the factors of $1/2$ appear due to the resetting transition from the second level to the third level (in $B\to C$ stroke) or to the ground level (in $D\to A$ stroke). 

\paragraph*{\textbf{Efficient Power:}} In heat engines, it is generally observed that the efficiency and output power are maximized under different conditions. For instance, a reversibly driven engine shows maximum efficiency, but the power output is zero due to the quasistatic nature of the driving. On the other hand, an engine operating at maximum power yields a smaller efficiency \cite{curzon1975efficiency}. A parameter that is often used to quantify an optimal engine is the product of its efficiency and output power, called the \textit{efficient power} (EP) \cite{Yilmaz2006,Singh2018,Lahiri2020}:
\begin{equation}
    {\mathrm{EP}} = \eta_{\mathrm{eff}}W_{\mathrm{out}}/(4\mathcal{T}).
    \label{eq:efficient power}
\end{equation}
If the EP assumes a high value in a region of the parameter space, it would imply that the engine has a reasonably high efficiency as well as power. Note that in the above definition, we have used $\eta_{\mathrm{eff}}$ instead of $\eta$, which is a more reasonable definition for appropriately  characterizing the quality of the engine.

\section{Results and Discussions}
\label{sec:results}
%\subsection{Two-level system as engine}

\paragraph*{\textbf{Two-level system as engine:}} We know that the state of the system in absence of resetting is given by $\rho^{\mathrm{nr}}(t)$, which is a solution of either Eq. (\ref{eq:von Neumann}) or (\ref{eq:Lindblad}), depending on whether or not the system is disconnected from a heat bath. This can be solved numerically, and the solution can be fed into the renewal equation given by Eq. (\ref{eq:renewal equation}) in order to obtain $\rho^{\mathrm{r}}(t)$. 

The solution can also be obtained by explicitly simulating the reset events, with Lindbladian evolution between two consecutive resets. In this case, after every evolution duration $\tau$ that is sampled from an exponential distribution $p(\tau)$ (see Eq. (\ref{eq:tau distribution})), a reset event occurs. In our simulations, instead of sampling $\tau$, we equivalently impose the dynamics given by Eq. (\ref{eq:reset dynamics}). With probability $rdt$, we force the system to suddenly switch to the reset state. For instance, if the two-level system is reset to a ground state at time $t=t'$, then we impose 
\[
\rho^{\mathrm{r}}(t') = 
\left(\begin{array}{cc}
    \rho_{11}(t') & \rho_{12}(t')\\
    \rho_{21}(t') & \rho_{22}(t')
\end{array}\right)
\to \rho^{\mathrm{reset}} = \rho_g = 
\left(\begin{array}{cc}
    0 & 0\\
    0 & 1
\end{array}\right).
\]
Since the reset events are random, an entire ensemble of realizations are required, over which the averaged thermodynamic quantities are to be computed.

\paragraph*{\textbf{Three-level system as engine:}} 
The protocol and the Lindblad operators for the three-level system have already been discussed in Sec. \ref{sec:thermodynamics}.  We consider a three-level system with consecutive energy levels separated by $\omega(t)/2$, where $\omega(t)$ is the energy gap used in the two-level system considered earlier. 
In other words, the energy eigenvalue of the excited state of the two-level system becomes that of the second excited state for the three-level system, while the first excited state of the latter appears midway between these two levels.
Let the energy gaps between the ground state and first excited state be $\omega_{01}$, and between the first and second excited states be $\omega_{12}$, where $\omega_{01} = \omega_{12} = \omega/2$. A suitable index $h$ or $c$  will be added to denote the energy gaps corresponding to strokes $B\to C$ and $D\to A$, respectively. For instance, 
when the system is connected to the hot bath, the frequency is given by $ \omega_{01}^h = \omega_{12}^h=\omega_h/2$. Similar definitions would be used for the cold bath.
 %The state of the system is reset to the second excited level in the $B\to C$ stroke and to the ground level in the $D\to A$ stroke,  in order to make an appropriate comparison with the two-level system. 
 The three energy levels that are available for the reset operations are the ground state $|g\rangle$, first excited state $|e_1\rangle$, and the second excited state $|e_2\rangle$, that correspond to the density matrices
 \begin{equation}
   \rho_g = \left(
   \begin{array}{ccc}
         0 & 0 & 0\\
         0 & 0 & 0\\
         0 & 0 & 1
     \end{array}\right);
     \hspace{1cm}
     \rho_{e_1} = \left(
     \begin{array}{ccc}
         0 & 0 & 0\\
         0 & 1 & 0\\
         0 & 0 & 0
     \end{array}\right);
     \hspace{1cm}
     \rho_{e_2} = \left(
     \begin{array}{ccc}
         1 & 0 & 0\\
         0 & 0 & 0\\
         0 & 0 & 0
     \end{array}\right).
 \end{equation}

 \paragraph*{\textbf{Comparison of analytical and simulation results:}} In order to proceed further, we note that for a given integration time step $\delta t$ used in the semi-analytical calculations involving the renewal equation (see Eq. (\ref{eq:renewal equation}) and the discussion below it), the error induced by approximating $(1-r\delta t)^n \simeq e^{-rt}$ increases with increase in $r$. This would lead to incorrect estimations of physical observables obtained from the numerical implementation of the renewal equation. Thus, to improve the accuracy of our semi-analytical computations, we keep the value of $r\delta t$ constant, so that an increase in the resetting rate automatically reduces the integration time step. We show in Fig. \ref{fig:comparison_rho} (see appendix \ref{sec:comparison}) that this is indeed the case for both the two- and three-level systems, by comparing the simulated and semi-analytical values of the population densities $\rho^{\mathrm{r}}(t)$ in states $|e\rangle$ and $|e_2\rangle$, respectively. 
 Henceforth we study our system using only the semi-analytical results. 
For the numerical integration of the renewal equation as well as of the Lindblad equations between two reset events, we have kept $\delta t = 0.01/r$.

\paragraph*{\textbf{Work extracted in different combinations of reset states:}}

\begin{figure}[!ht]
    \centering
    \begin{subfigure}{0.48\linewidth}
        \includegraphics[width=\textwidth]{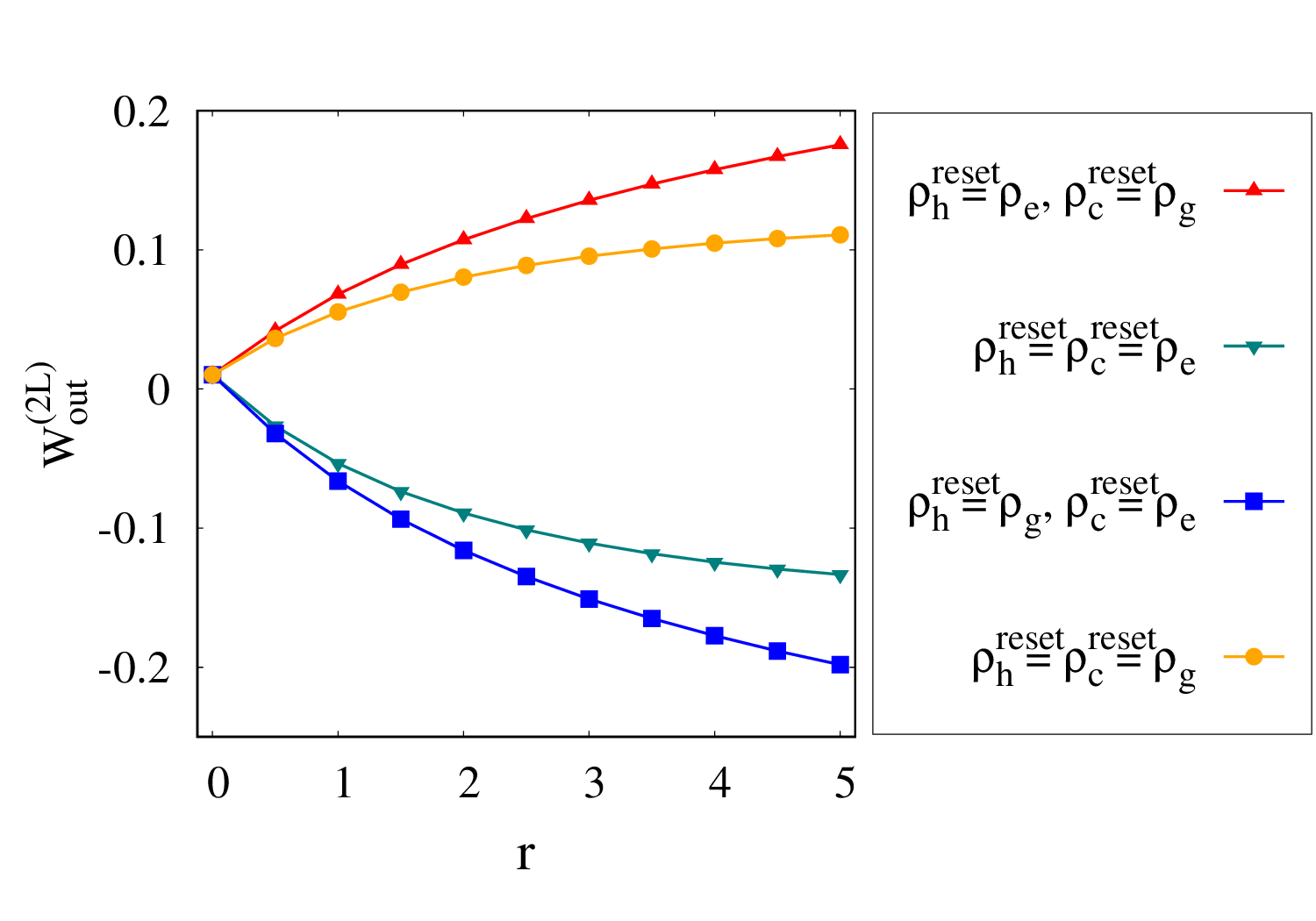}
        \caption{}
    \end{subfigure}
    \hfill
    \begin{subfigure}{0.48\linewidth}
        \includegraphics[width=\textwidth]{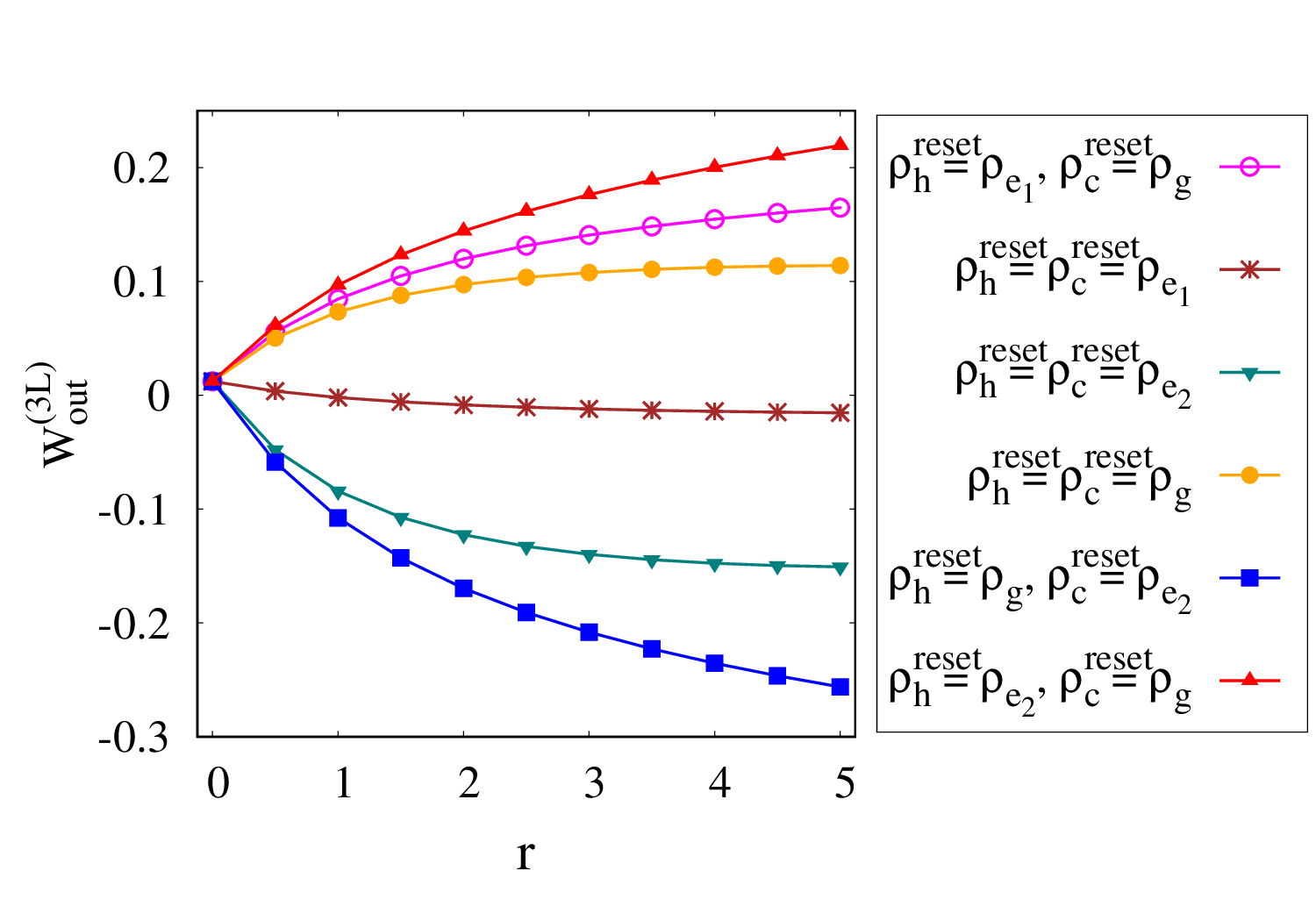}
        \caption{}
    \end{subfigure}
    \caption{(a) Comparison between $W_{\mathrm{out}}$ for different pairs of resetting states for the hot and the cold baths, for the two-level engine. At $r = 0$, the value of the output work is $W_{\mathrm{out}}^{(2L)}\simeq 0.020$. The parameters used are $\omega_h = 1$, $\omega_c = 0.5$, $T_h = 4$, $T_c = 1.5$, $\delta t = 0.01/r$, and $\mathcal{T} = 2$.  (b) Similar comparison for the three-level engine. At $r = 0$, the value of the output work is $W_{\mathrm{out}}^{(3L)}\simeq 0.024$. The parameters used are $\omega_{01} = \omega_{12} = \omega/2$, and $\omega_{13} = \omega_{31} = \omega$, where $\omega$ equals to either $\omega_c$ or $\omega_h$, corresponding to the cold and hot baths, respectively. Other parameters are the same as for the two-level engine.}
    \label{fig:work_output_vs_rr_diff_reset_state}
\end{figure}

 In Fig. \ref{fig:work_output_vs_rr_diff_reset_state}(a), we have plotted the analytical values of extracted work for the two-level engine as a function of time, for different combinations of reset states $\rho^{\mathrm{reset}}$ in hot and cold baths (referred to as $\rho_h^{\mathrm{reset}}$ and $\rho_c^{\mathrm{reset}}$ respectively), in order to check which of these combinations yields the maximum output. These combinations are: \vspace{-0.2cm}\\
  \begin{enumerate}\itemsep -0.05cm
    \item $\rho_h^{\mathrm{reset}} = \rho_g$ and $\rho_c^{\mathrm{reset}} = \rho_e$ (blue curve with solid squares)
  
     \item $\rho_h^{\mathrm{reset}} = \rho_c^{\mathrm{reset}} = \rho_e$ (green curve with solid downward triangle)

     \item $\rho_h^{\mathrm{reset}} = \rho_c^{\mathrm{reset}} = \rho_g$ (orange curve with solid circles)

     \item $\rho_h^{\mathrm{reset}} = \rho_e$ and $\rho_c^{\mathrm{reset}} = \rho_g$ (red curve with solid upward triangle)
 \end{enumerate}

 It is readily observed that the extracted work is higher in the fourth case. The reason can be eluciated by the following argument. In absence of resets, the decrease in energy of the more populated ground state during the expansion stroke $A\to B$ leads to work being extracted, while work must be injected into the system in the compression stroke $C\to D$. Now, since in stroke $B\to C$, the hot bath increases the relative population of the excited state, the work extracted exceeds the work injected, so as to make the cyclic protocol work as an engine. This effect gets more pronounced when the resetting condition further increases the relative population of the excited state in the $B\to C$ stroke, and conversely decreases it during the $D\to A$ stroke. This is accomplished by condition (iv) above. For this reason, we would adhere to this protocol throughout this manuscript. 

 In Fig. \ref{fig:work_output_vs_rr_diff_reset_state}(b), we carry out similar study for the three-level system. The general trends of work as a function of $r$ remain similar to that of the two-level system, for different combinations of the reset states in the two heat baths. 
 However, there are two additional curves where the reset state combinations are $\rho_h^{\mathrm{reset}} = \rho_{e_1}, ~\rho_c^{\mathrm{reset}} = \rho_g$ (magenta line with open circles) and $\rho_h^{\mathrm{reset}} = \rho_c^{\mathrm{reset}} = \rho_{e_1}$ (brown line with stars), respectively.
 We find that the combination $\rho_h^{\mathrm{reset}}=\rho_{e_2}$ and $\rho_c^{\mathrm{reset}}=\rho_{g}$ yields the maximum work output of the engine. Furthermore, in comparison with Fig. \ref{fig:work_output_vs_rr_diff_reset_state}(a), we find that the values of extracted work are higher for the 3-level system, even though the difference between the states $|g\rangle$ and $|e_2\rangle$ remains the same. This is because the intermediate level now allows a lower fraction of particles to be present in $|e_2\rangle$, which is a property of the Boltzmann factor.

\paragraph*{\textbf{Output work as a function of the reset rate:}}
    As shown in Fig. \ref{fig:work_output_vs_rr_diff_reset_state}, the extracted works $W^{(2L)}_{\mathrm{out}}$ and $W^{(3L)}_{\mathrm{out}}$ (the superscripts indicate a two-level or a three-level system) are observed to increase as a function of the resetting rate, provided the state is reset to an excited state in the $B\to C$ stroke and to the ground state in the $D\to A$ stroke. Consequently, the engine performs better in terms of output power, as the value of $r$ increases.

 \begin{figure}[!ht]
 \centering
 	\includegraphics[width=0.5\linewidth]{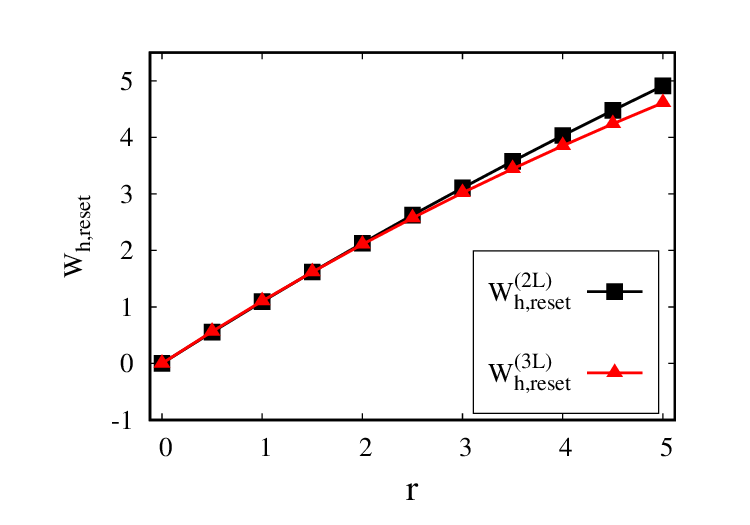}
      \caption{Plot showing $W_{h,{\mathrm{reset}}}$, as a function of $r$ for two- and three-level systems. Here, $\rho_h^{\mathrm{reset}}=\rho_e$, $\rho_c^{\mathrm{reset}} = \rho_g$ for the two-level and $\rho_h^{\mathrm{reset}}=\rho_{e_2}$, $\rho_c^{\mathrm{reset}} = \rho_g$ for the three-level system. Other parameters used are the same as in Fig. \ref{fig:work_output_vs_rr_diff_reset_state}.}
    \label{fig:ana_simul_cost_of_resetting_rr}
 \end{figure}

\paragraph*{\textbf{Cost of resetting:}} 
Fig. \ref{fig:ana_simul_cost_of_resetting_rr} shows the dependence of the cost of resetting (minimum average input energy required for sustaining a certain resetting rate) $W_{h,{\mathrm{reset}}}$ for the $B\to C$ stroke
 for the two-level (black curve with solid squares, labelled by the superscript $(2L)$) and the three-level (red curve with solid triangles, labelled by the superscript $(3L)$) engines (see Eq. (\ref{eq:W reset})), on the resetting rate $r$. The reset states for the hot and cold baths are as discussed above (also mentioned in the figure caption). We observe that $W_{h,{\mathrm{reset}}}$ increases with $r$, as expected. Further, its magnitude is much higher than the work extracted from the engine, as can be seen by comparing with the red curves in Fig. \ref{fig:work_output_vs_rr_diff_reset_state}.

 \begin{figure}[!ht]
 \centering
 \begin{subfigure}{0.48\linewidth}
        \includegraphics[width=\textwidth]{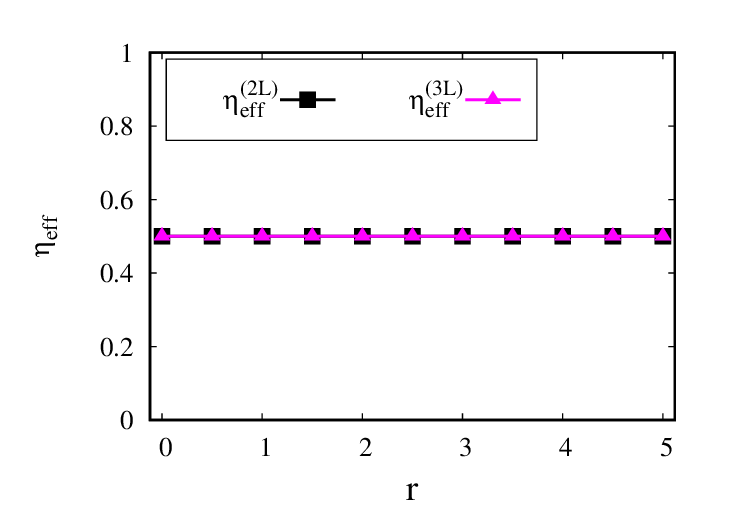}
        \caption{}
    \end{subfigure}
    \begin{subfigure}{0.48\linewidth}
        \includegraphics[width=\textwidth]{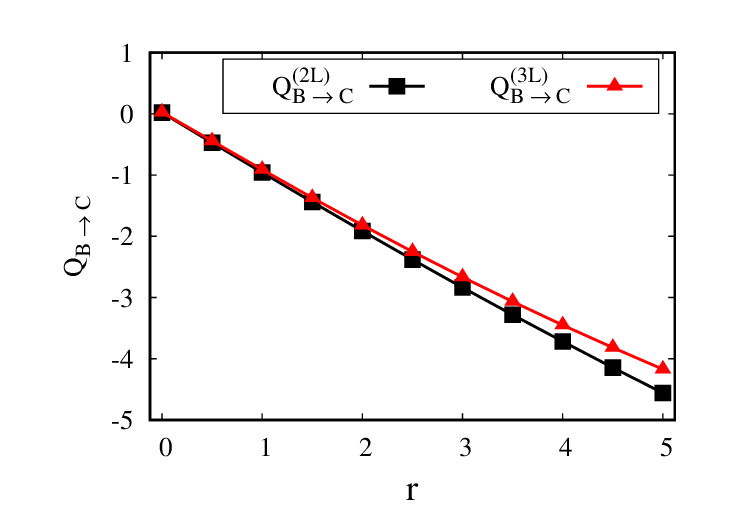}
        \caption{}
    \end{subfigure}
      \caption{Plot shows the $Q_{B\to C}$  as the function of $r$ for the two-level and three-level engines. The parameters used are the same as in Fig. \ref{fig:work_output_vs_rr_diff_reset_state}.}
     \label{fig:Qbc_efficiency}
 \end{figure}

\paragraph*{\textbf{Efficiency:}} 
The value of $\eta_{\mathrm{eff}}$ is found to be a constant, equalling $(1-\omega_c/\omega_h)$, for both the two-level and three-level Otto engines, just like the no-reset case (see Appendix \ref{sec:quasistatic_efficiency} and \cite{kumar2023thermodynamics} for details). This has been shown in Fig. \ref{fig:Qbc_efficiency}(a).  However, the value of $\eta$ turns out to be negative for the set of parameters used here. This happens because the value of $Q_{B\to C} = \tilde Q_{B\to C}-W_{h,{\mathrm{reset}}}$ becomes negative, even though $\tilde Q_{B\to C}$ is positive. 
In fig. \ref{fig:Qbc_efficiency}(b), we plot $Q_{B\to C}$ as a function of $r$ for the two-level and the three-level engines. The values are found to become more negative with increasing $r$.
The physical reason is the increased relative population of the excited state during the $B\to C$ stroke due to resetting. This, counter-intuitively, causes heat to be released in place of being absorbed, even though the system is in contact with the hot bath. 
 Thus, the value of $\eta$ gives no indication of the fact that the system is actually working as an engine for the chosen set of parameters.

  \begin{figure}[!ht]
 \centering
 	\includegraphics[width=0.5\linewidth]{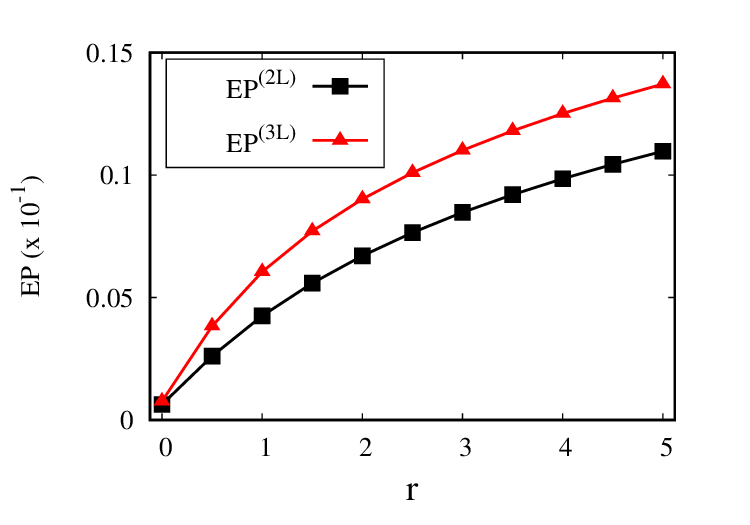}
      \caption{Plots showing ${\mathrm{EP}}$ as a function of $r$ for the two- and three- level engines, denoted by ${\mathrm{EP}}^{(2L)}$ and ${\mathrm{EP}}^{(3L)}$, respectively. The parameters used are the same as in Fig. \ref{fig:work_output_vs_rr_diff_reset_state}.}
     \label{fig:efficient power}
 \end{figure}

 \paragraph*{\textbf{Efficient power:}} Next, we study the variance of the efficient power, EP (see Eq. (\ref{eq:efficient power})), as a function of the resetting rate. The plots are shown in Fig. \ref{fig:efficient power}, for the two-level (denoted by ${\mathrm{EP}}^{(2L)}$) and three-level (denoted by ${\mathrm{EP}}^{(3L)}$) engines, for the same set of parameters. 
 The values are observed to increase monotonically with reset rate.
 The overall trends for the two- and three-level systems match those of $W_{\mathrm{out}}^{(2L)}$ and $W_{\mathrm{out}}^{(3L)}$, given by the red curves in Figs. \ref{fig:work_output_vs_rr_diff_reset_state}(a) and (b), since $\eta_{\mathrm{eff}}$ is a constant for both systems.
 Clearly, defining the ${\mathrm{EP}}$ to be a product of the $\eta$ (see Eq. (\ref{eq:eta})) and the output power would yield a negative value of EP, since $\eta$ becomes negative (see the discussions on Fig. \ref{fig:Qbc_efficiency}). Thus, we again find that the effective efficiency more reliably evaluates the engine's performance.

 \section{Thermodynamic work and reset work} \label{sec:thermodynamic_reset_work}

 Incorrectly considering the reset work as a part of the thermodynamic work can lead to erroneous conclusions. The problem becomes more subtle when the engine does not have a clear separation of the heat stroke and the work stroke, like the Carnot or Stirling cycle. In these cases, the same stroke can allow the system to exchange heat with the bath as well as do work. Consider the no-reset case. The net change in internal energy in one of these strokes (say the one where the system is placed in contact with the hot bath), will be given by
 \begin{equation}
    \langle\Delta E\rangle_{h} = \Delta{\mathrm{Tr}}[\rho(t)H(t)] = \int_\mathcal{T}^{2\mathcal{T}} dt~{\mathrm{Tr}}\left[\frac{\partial\rho(t)}{\partial t}H(t)\right] + \int_\mathcal{T}^{2\mathcal{T}} dt~{\mathrm{Tr}}\left[\rho(t)\frac{\partial H(t)}{\partial t}\right].
 \end{equation}
 In standard quantum thermodynamics, the first term on the right-hand-side is identified with the heat absorbed in that stroke, while the second term is the \textit{thermodynamic work} done on the system. However, in our case, because of the presence of resetting, the thermodynamic work has to be differentiated from the resetting work (see the formalism of \cite{lahiri2024efficiency}). The net work done on average in a given time step $\delta t$ in this stroke must be a sum of the resetting work (which takes place with a probability of $r\delta t$) and the thermodynamic work (which takes with a proability of $(1-r\delta t)$). Thus, the increment in total work on the system in a given time step will be
 \begin{equation}
     W_{h,{\mathrm{tot}}}(t+dt) = W_{\mathrm{tot}}(t) + r\delta t(E^{\mathrm{reset}} - \langle E(t)\rangle) + (1-r\delta t){\mathrm{Tr}}\left[\rho^{\mathrm{r}}(t)\frac{\partial H(t)}{\partial t}\delta t\right].
 \end{equation}
 Here, $E^{\mathrm{reset}} = {\mathrm{Tr}}[\rho^{\mathrm{reset}}H(t)]$ is the expectation value of energy in the state to which the reset is taking place. Rearranging the terms, for the infinitesimal time step limit $\delta t \to dt$ and retaining terms up to first order in $dt$ on the right-hand side, we obtain $dW_{\mathrm{tot}}/dt$, which can be integrated to yield
 \begin{equation}
     W_{h,{\mathrm{tot}}}(t) =  r\int_\mathcal{T}^{2\mathcal{T}} dt~(E^{\mathrm{reset}} - \langle E(t)\rangle) + \int_\mathcal{T}^{2\mathcal{T}} dt~{\mathrm{Tr}}\left[\rho^{\mathrm{r}}(t)\frac{\partial H(t)}{\partial t}\right].
 \end{equation}
  The first term on the right-hand side is the reset work $W_{h,{\mathrm{reset}}}$ during this stroke, while second term is the thermodynamic work. As per our sign conventions, the negative of the second term gives the output work. The contribution of the first term is already taken into account when we compute $\tilde Q_{B\to C}$. In a similar manner, $W_{c,{\mathrm{tot}}}$ can be defined. If $W_{h,{\mathrm{tot}}}$ and $W_{c,{\mathrm{tot}}}$ are used while computing $\eta_{\mathrm{eff}}$ (see Eq. (\ref{eq:eta effective})) instead of the corresponding thermodynamic works, then the contributions of $W_{h,{\mathrm{reset}}}$ and $W_{c,{\mathrm{reset}}}$ appear both in the numerator and the denominator, thereby leading to incorrect results.

\section{Conclusions}
\label{sec:conclusions}

In this work, we have studied the effect of stochastic resetting of a system's energy state on the engine's output work and efficiency. We have used a two-level system as well as a three-level system as engine, and have compared the above observables for these two cases. The Lindblad master equation has been used to obtain the results. In our simulations, the process of resetting has been added and an ensemble of realizations are generated in order to compute the additional averaging due to the random reset processes. For the analytical results, we numerically extract the solution of the Lindblad equation in the absence of resetting (pure thermal baths), and use the renewal equation to obtain results for the case where resetting is present. We have defined an effective efficiency $\eta_{\mathrm{eff}}$ that by definition contains the input energy provided to maintain a given resetting rate in the steady state. This has been contrasted with the standard definition of efficiency, where the energy input from reset operations have been ignored. The effective efficiency is shown to be the more meaningful quantifier for the engine's capacity to generate useful work. We have also studied the behavior of the product of output power and $\eta_{\mathrm{eff}}$, namely, the efficient power, as a function of the resetting rate. This parameter is considered to be a robust indicator of the usefulness of the engine, as compared to efficiency or output power. Unlike the effective efficiency that remains constant with change in resetting rate, the efficient power increases with $r$. The three-level system is observed to yield higher output work and efficient power as compared to the two-level system. An interesting extension of this work would be to study the effect of implementing a probabilistic and finite-time resetting. 

\appendix
\section{Comparison between analytical and simulation results}
\label{sec:comparison}

\begin{figure}[!ht]
    \centering
    \begin{subfigure}[t]{0.48\linewidth}
        \centering
        \includegraphics[width=\linewidth]{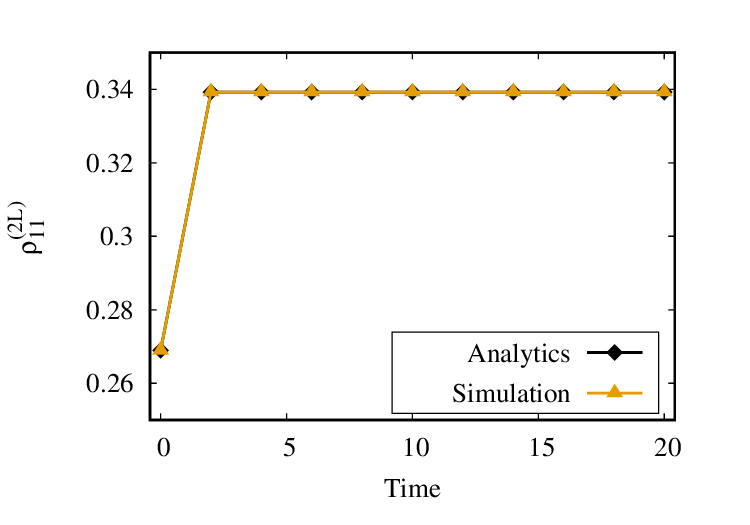}
        \caption{}
           \end{subfigure}
    \hfill
    \begin{subfigure}[t]{0.48\linewidth}
        \centering
        \includegraphics[width=\linewidth]{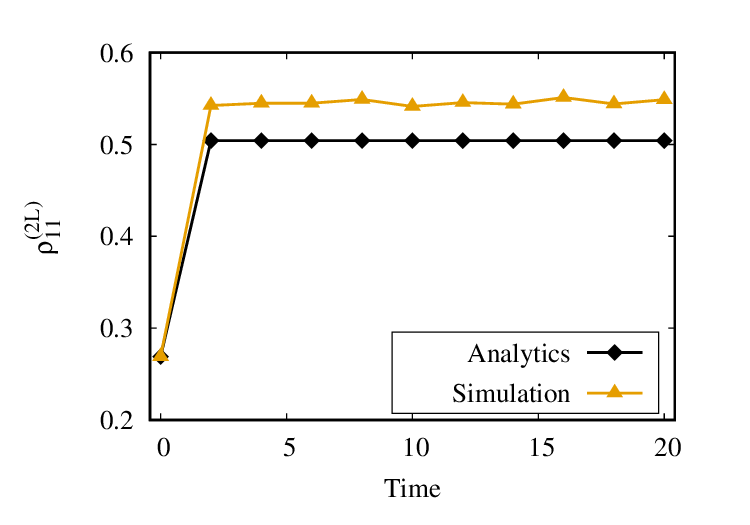}
        \caption{}
           \end{subfigure}
    \vfill
    \begin{subfigure}[t]{0.48\linewidth}
        \centering
        \includegraphics[width=\linewidth]{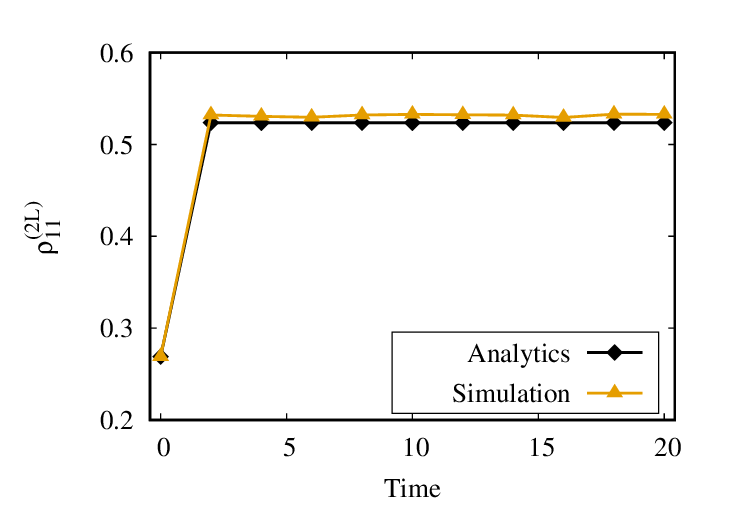}
        \caption{}
         \end{subfigure}
         \begin{subfigure}[t]{0.48\linewidth}
             \centering\includegraphics[width=\linewidth]{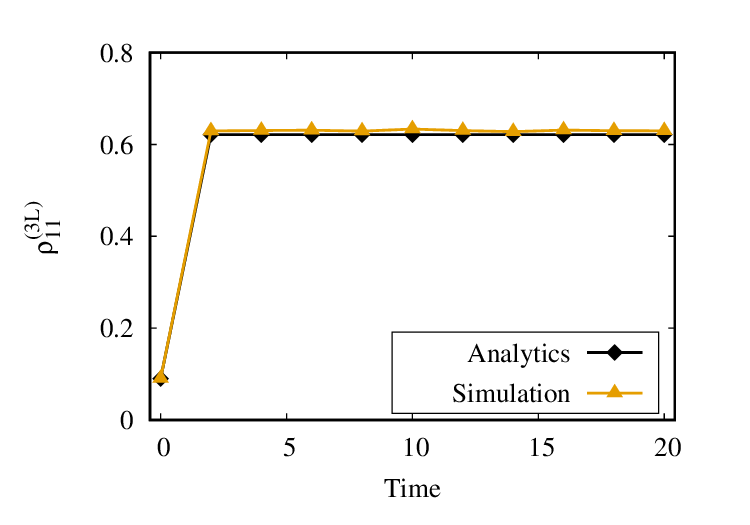}
             \caption{}
         \end{subfigure}
    \caption{The figure presents a comparison between analytics and simulations for the population densities. For the two-level system, the different cases considered are (a) $\delta t = 0.01$ with $r=0$, (b) $\delta t = 0.01$ with $r=5$ , and (c) $\delta t=0.01/r$ with $r=5$. The parameters are: $~\omega_h = 10 $, $~\omega_c = 5 $, $~T = 15 $, $\gamma_0=1$, $~\mathcal{T} = 2 $, and $~\rho^{\mathrm{reset}} = \rho_e$. In (d), we present a comparison for the three-level system with $\delta t=0.01/r$ with $r=5$. Here,  $\rho^{\mathrm{reset}} = \rho_{e_2} $. Other parameters are the same as for the two-level system.}
    \label{fig:comparison_rho}
\end{figure}
In Fig. \ref{fig:comparison_rho}, we compare the simulation and analytical results for a two-level system with energy gap $\omega$ in contact with a heat bath at temperature $T$ for three cases: (a) $r=0$ (purely thermal evolution), (b) $\delta t=$constant, and (c) $r\delta t=$ constant.
  We observe that even though case (a) shows a good agreement between analytics and simulations, it ceases to show it when resetting is implemented, as is evident from (b). However, keeping $r\delta t$ constant reduces the discrepancy with the simulation results, as observed from (c). This agreement establishes the authenticity of our codes used for both semi-analytics and simulations.  In (d), we provide a similar comparison for a three-level system when $r\delta t$ is held fixed. It has been separately verified that keeping $\delta t$ fixed decreases the accuracy for the three-level system as well, as per our expectations.

\section{Thermodynamic quantities in the quasistatic regime without resetting}
\label{sec:quasistatic_efficiency}
If the engine cycle is carried out very slowly (quasistatically) in absence of reset, the density matrix at the states at each corner of Otto cycle A, B, C and D would correspond to the Boltzmann distribution. For the two-level system, the average energy values at these points become
\begin{eqnarray}
    \langle E_A \rangle &=& -\omega_c \tanh(\beta_c \omega_c); \nonumber \\
    \langle E_B \rangle &=& -\omega_h \tanh(\beta_c \omega_c); \nonumber \\
    \langle E_C \rangle &=& -\omega_h \tanh(\beta_h \omega_h); \nonumber \\
    \langle E_D \rangle &=& -\omega_c \tanh(\beta_h \omega_h).
    \label{eq:avg_energy}
\end{eqnarray}

Using the expressions for average energy in Eq.~(\ref{eq:avg_energy}), and using the definitions of $W_{\mathrm{out}}$  and $Q_{B\to C}$ as given in Sec. \ref{sec:thermodynamics}, we obtain the expression for efficiency to be $(1-\omega_c/\omega_h)$.

For the three-level system, the expressions for average energies are lengthier, but the expression for efficiency is identical to that of the two-level system.

\newpage
%\bibliographystyle{vancouver}
%\bibliography{ref}

\end{document}